\documentclass[aps,pre,twocolumn,float]{revtex4}
\usepackage{amsmath,bm,epsfig}
\newcommand{\ud}{\mathrm{d}}

\newcommand{\B}[1]{{\bm{#1}}}%% Bold Roman & Greek Lower & Upper Case
\begin{document}
%%%%%%%%%%%%%%%%%%%%%%

\title{Mechanical Properties and Plasticity of a Model Glass Loaded Under Stress Control}
\author{Vladimir Dailidonis$^{1,2}$,  Valery Ilyin$^1$,  Pankaj Mishra$^1$ and Itamar Procaccia$^1$}
\affiliation{$^1$Weizmann Institute of Science,  Rehovot 76100, Israel \\
$^2$Bogolyubov Institute for Theoretical Physics, 03680 Kiev, Ukraine}
\begin{abstract}
Much of the progress achieved in understanding plasticity and failure in amorphous solids had been
achieved using experiments and simulations in which the materials were loaded using strain control.
There is paucity of results under stress control. Here we present a new method that was carefully
geared to allow loading under stress control either at $T=0$ or at any other temperature, using
Monte-Carlo techniques. The method is applied to a model perfect crystalline solid, to a crystalline solid
contaminated with topological defects, and to a generic glass.  The highest yield stress belongs to the crystal, the lowest to the crystal with a few defects, with the glass in between. Although the glass is more disordered than the crystal with a few defects, it yields stress is much higher than that of the latter. We explain this fact by considering the actual microscopic interactions that are typical to glass forming materials, pointing out the reasons for the higher cohesive nature of the glass. The main conclusion of this paper is that the instabilities
encountered in stress-control condition are the identical saddle-node bifurcation seen in strain-control. Accordingly one can use the latter condition to infer about the former. Finally we discuss temperature effects and comment on the time needed
to see a stress controlled material failure.

\end{abstract}
\maketitle
%%%%%%%%%%%%%%%%%
\section{Introduction}

Plasticity in crystalline solids is known to be carried by defects, typically dislocations, that glide irreversibly
under the influence of loading the material with some mechanical load \cite{MS73,C53}. On the other hand, the study of plasticity and
yield in amorphous solids is an ongoing subject of research, with many issues remaining to be discovered, especially in
more complex amorphous glasses like polymeric glasses and metallic glasses. Much of the recent progress in understanding
plasticity in amorphous solids was based on experiments and simulations done by loading the system under strain control
protocols. A useful simulational protocol that attracted much attention is the quasi-static athermal (AQS) strain control protocol,
in which the system is maintained at zero temperature, and is allowed to return to mechanical equilibrium after every small increase in strain
\cite{ML04,KLLP10}.
This protocol exposed very nicely the role of mechanical instabilities. These are easily detected by examining the Hessian matrix of the system;
the eigenvalues of this matrix are all positive when the system is mechanically stable, while a plastic instability is characterized by an eigenvalue
approaching zero, typically via a saddle-node bifurcation \cite{98ML}. When this happens, the associated eigenfunction, which is also identified with the non-affine
response of the system, localizes on a sub-set of particles, those that participate in the plastic event.

In this paper we examine the corresponding physics for stress controlled protocols. In some sense, this is the more
natural protocol because it provides one with the right control to precisely determine when does the system yield
in the sense that its strain will increase indefinitely as long as the stress is maintained at a given value.
When the stress is below the yield stress $\sigma_{_{\rm Y}}$ the strain will reach a limit. Indeed, some attempts
to study yield using stress controlled simulations were reported in the literature \cite{FL98,RS09}. We propose
a more straightforward protocol that appears to provide highly stable results which are in good correspondence
with the best available strain-controlled results.
The new protocol is introduced in Sect. \ref{protocol}. In Sect. \ref{model} we present the physical models employed here. We
discuss stress-controlled loading of a perfect hexagonal structure in 2-dimensions, the same structure marred by
some defects, and finally a generic binary glass. This section includes some of the main conclusions of this
study: we argue that the instabilities seen in stress control loading are the very same saddle-node bifurcations that
are exhibited in strain-controlled experiments. The difference is that once the system yields in stress-control there is
no recovery. In strain controlled loading the system can yield, release a portion of its stress, and then be loaded
again, to yield again etc. Therefore one sees the typical serrated stress vs. strain curves that can go for some time
up to high values of the strain. In contrast, in stress controlled experiments the system either gets stuck if the applied stress is smaller than the yield stress, or it fails if the stress is higher than the yield stress. We show that the knowledge of strain controlled results is useful in predicting much of what can happen in stress controlled loading.
In Sect. \ref{temp} we focus on thermal effects, and particularly what happens when the stress is lower than the
yield stress but temperature fluctuations can result in surmounting the barrier and failing. Predicting the waiting time
becomes an easy exercise once one realizes that the transition is due to a saddle node bifurcation. This fact implies that the eigenvalue that vanishes at the transition has a square-root singularity, and together with the generic dependence of the barrier hight on the distance from the bifurcation one can easily estimate the waiting time.
Sect. \ref{summary} offers a summary and some concluding remarks.

\section{Statistical mechanics of loaded systems}
\label{protocol}

In this section we construct a protocol based on a method that was proposed for the simulations of deformations in solids in Ref. \cite{PR80}. The main ingredient in this approach is in changing the shape of
the simulation box as well as its size. In principle this approach can be adapted to either molecular dynamics or Monte Carlo techniques as can be see in e.g. Refs. \cite{PR80,PR81,WTK03}). This method can be used even for large deformations under applied external forces, see Refs. \cite{RR84,LB94}.

In the variable shape method \cite{PR80,PR81} the particle positions change from the reference state $\{\B r^0_i\}$ to a new one,
denoted $\{\B r_i\}$, by an affine transformation that is defined by a matrix ${\bf J}$ :
\begin{equation}
\B r_i=\B J\cdot\B r_i^0.
\label{ATJ}
\end{equation}

On the microscopic level the affine transformation Eq.~(\ref{ATJ}) destroys mechanical equilibrium, and it should be followed by a non-affine atomic-scale relaxation of the particle positions $\{\B r^0_i\}$ \cite{LM06}.
This relaxation can be performed by Molecular Dynamics or Monte Carlo methods or in the case of a athermal system by energy minimization.

In the frame of statistical mechanics the mean value of an observable
in a loaded system is defined by
\begin{equation}
\langle A\rangle =\frac{\int {\rm d} \B J {\rm d} \B r_1^0 \cdots {\rm d}\B r_N^0~ A(\B  r_i^0,\B J)\cdot
e^{-G(\{\B  r_i^0\},\B J,\B \sigma^{ext})/T}}{\int \ud \B J\cdot  \B r_1^0 \cdots {\rm d}\B r_N^0
\cdot e^{-G(\{\B  r_i^0\},\B J,\B\sigma^{ext})/T}},
\label{avSM}
\end{equation}
Here $T$ is the temperature and $G(\{\bf r_i^0\},\B J,\B \sigma^{ext})$ is the generalized enthalpy
and $\B \sigma^{ext}$ is
the external stress tensor.   The Monte Carlo method allows to evaluate this
expression numerically.

The method of variable shape introduces strain into the simulation box by first defining a square box of {\em unit area} where the particles are at positions $\B s_i$. Next one defines a linear transformation $\B h$, taking the particles to positions $\B r_i$ via $\B r_i={\bf h}\cdot\B s_i$. In order to prevent rotations of the simulation box, the matrix ${\bf h}$ should be symmetric. The current area of a system becomes the determinant
$V=\mid {\bf h}\mid$.
Then the positions of the particles in the reference state are defined by $\B r^0_i={\bf h}_0\cdot\vec{s}_i$;
accordingly the matrix ${\bf J}$ in Eq.~(\ref{ATJ}) is given by ${\bf J}={\bf h}\cdot{\bf h}_0^{-1}$.

It is suitable to change integrals over the  components of the matrix $\B J$  in Eq.~(\ref{avSM}) by integrals over the independent components of the matrix ${\bf h}$ and the integrals over ${\bf r_i^0}$ by integrals over ${\bf S}=\{\vec{s}_i\}$. Then this equation reads
\begin{equation}
\langle A\rangle =\frac{\int \ud {\bf h}\cdot \ud {\bf S}\cdot \cdot A({\bf S},{\bf h})\cdot
e^{-G^{\prime}({\bf S},{\bf h},\boldsymbol\sigma^{ext})/T}}{\int \ud {\bf h}\cdot \ud {\bf S}\cdot
\cdot e^{-G^{\prime}({\bf S},{\bf h},\boldsymbol\sigma^{ext})/T}},
\label{avH}
\end{equation}
where
\begin{equation}
G^{\prime}({\bf S},{\bf h},\boldsymbol\sigma^{ext})
=-T N\ln V +
 G({\bf S},{\bf h},\boldsymbol\sigma^{ext}).
\label{Gpr}
\end{equation}

The integral in Eq.~(\ref{avH}) is evaluated via the Metropolis algorithm.
Two kinds of trial moves are considered: one performs $n$ standard Monte Carlo moves (displacement of the particle positions given by $\vec{s}_i$)
\begin{equation}
\B s^{new}_i=\B s^{old}_i+\delta \B s,\hspace{4 mm} 1\le i\le N.
\label{Rmove}
\end{equation}
In this equation the $\alpha$ component of the displacement vector of a particle is given by
\begin{equation}
\delta s^{\alpha}=\Delta s_{max}(2\xi^\alpha-1),
\label{ParDisp}
\end{equation}
where $\Delta s_{max}$ is the maximum displacement and
$\xi^\alpha$ is an independent random number uniformly distributed between 0 and 1
After $n$ sweeps defined by Eq.~(\ref{Rmove})
 the transformation ${\bf h}$ changes according to
\begin{equation}
\B h^{new}=\B h^{old}+\delta\B h,
\label{hnew}
\end{equation}
where elements of the random symmetric matrix $\delta\B h$ are defined by
\begin{equation}
\delta h_{ij}=\Delta h_{max}(2\xi_{ij}-1),\hspace{4mm} i\le j.
\label{transH}
\end{equation}
Here $\Delta h_{max}$ is the maximum allowed change of a matrix element and
$\xi_{ij}$ is an independent random number uniformly distributed between 0 and 1.
The value of $\Delta h_{max}$  and the maximum displacement of particle
positions $\Delta s_{max}$ are selected so that the acceptance rate is $30\%$.
For each  kind of move the trial configuration is accepted with probability
\begin{equation}
P_{tr}=\min \bigg[1,exp\bigg(-\frac{\Delta G^{\prime}}{T}\bigg)\bigg].
\label{trial}
\end{equation}
For relaxation of particle positions the matrix $\B h$ is fixed and the difference of the
generalized enthalpy is defined by the difference of the potential energy of the system
$U(\B h,\{\B s\})$
\begin{eqnarray}
\Delta G^{\prime}&=&U(\B h,\B s_1,\cdots,\B s_i^{new},\cdots \B s_{N})- \nonumber \\
& &U(\B h,\B s_1,\cdots,\B s_i^{old},\cdots \B s_{N}),\hspace{4 mm} 1\le i\le N.
\label{DifU}
\end{eqnarray}
The change of the generalized enthalpy due to affine transformation (at fixed particle positions
$\{\B s\}$) is given by
\begin{eqnarray}
\Delta G^{\prime}&=&-T N\ln \big(V^{new}/V^{old}\big) +U(\B h^{new},\{\B s\})\nonumber \\
&-&U(\B h^{old},\{\B s\})+\delta W,
\label{difH}
\end{eqnarray}
where $\delta W$ s the work that is done by an external stress $\B \sigma^{ext}$.
In general case for move $\B J\to \B J+\delta \B J$ this work is given by (see, e.g., \cite{WLY95}
\begin{equation}
\delta W=-\frac{1}{2}V^{old}Tr\big( \B \sigma^{ext}(\delta \B J {\B J}^{-1}+\tilde{\B J}^{-1}\delta\tilde{\B J})\big).
\label{Work}
\end{equation}
Here the symbol $\tilde~$ represents the transpose of a matrix.
Taking into account the relation between the matrices $\B J$ and $\B h$ this equation can be written as
\begin{equation}
\delta W=-\frac{1}{2}Tr\big( \B \sigma^{ext}(\delta \B h {\B H}+\tilde{\B H}\delta\tilde{\B h})\big),
\label{WorkH}
\end{equation}
where the matrix $\B H$ is given by
\begin{equation}
\B H=\left(
\begin{array}{cc}
h_{yy}&-h_{xy}\\
-h_{xy}&h_{xx}
\end{array}
\right)
\label{MatrH}
\end{equation}

It follows from Eq.~(\ref{trial}) that in the limit $T\to 0$ only the
configurations with decreasing enthalpy are accepted, i.e., the Monte Carlo
process converges to one configuration with minimal generalized enthalpy (for $T=0$ the generalized
enthalpy is equal to the Gibbs free energy).
In general, this configuration belongs to a local minimum of the generalized enthalpy landscape and its position
depends on the initial configuration of the simulation process.

To specialize the technique described above to stress-controlled simple shear simulations at
zero temperature one chooses the
following ${\bf h}$ matrix
\begin{equation}
{\bf h}=L\left(
\begin{array}{c c}
1&\gamma \\
0&1
\end{array}
\right),
\label{hG}
\end{equation}
where $L$ is the length of the square simulation box and $\gamma$ is the simple shear strain, with the volume of the system V=$L^2$ being conserved. The external stress in this protocol is given by
\begin{equation}
\B \sigma=\left(
\begin{array}{c c}
0&\sigma^{ext}_{xy} \\
\sigma^{ext}_{xy}&0
\end{array}
\right) \ .
\label{ExtStr}
\end{equation}
For the matrix $\B h$ defined by Eq.~(\ref{hG}) the change of the generalized enthalpy due to the increment
 $\delta \gamma$ is  given by
\begin{equation}
\Delta G^{\prime}=U(\B \gamma+\B {\delta \gamma},\B r_i^{new})-
U(\B \gamma,\B r_i^{old})-V\sigma^{ext}_{xy}\delta\gamma.
\label{entG}
\end{equation}

Summing Eq.~(\ref{entG}) over infinitesimal increments one can find the generalized enthalpy
at a given state parameterized by $\gamma$, relative to the state defined by $\gamma_0$

\begin{equation}
G^{\prime}(\gamma,\gamma_0,\sigma^{ext}_{xy})=U(\B \gamma,\B r_i^{\gamma})-
U(\B \gamma_0,\B r_i^{\gamma_0})-V\sigma^{ext}_{xy}(\gamma-\gamma_0),
\label{entGT}
\end{equation}
where $U(\B \gamma,\B r_i^{\gamma})$ is the energy that is achieved after a sequence of steps in the frame of
this protocol. Usually the reference state corresponding  to $\gamma_0$ is defined at $\sigma^{ext}_{xy}=0$.
Nevertheless, as one can see from Eq.~(\ref{entGT}) the replacement of the reference state generates only a shift by a constant in
the generalized enthalpy; Once the generalized enthalpy is minimized the location of the minima do not depend on the reference state.

Note that the strain $\gamma$ appears explicitly in our formalism. It is therefore important to stress that
in general the strain {\em is not} a state function if the system undergoes irreversible
events during the nonaffine position reshuffling in which energy can be lost to the heat bath \cite{01Gil}. The generalized enthalpy
is determined by $\gamma$ as a state function only in the case of pure elasticity. Here the appearance of $\gamma$ in the
formalism should be interpreted only as a marker to the present shape of the system, and the energy has to be computed
incrementally via following the protocol.

In the next section we present the results of MC calculations for the
temperature $T=0.05$ and the pressure $P=0$ at different values of applied
shear stress. For the sake of more easy interpretation these results are
compared with the consequences of the AQS strain-controlled protocol.

%%%%%%%%%%%%%%%%
\section{The model and simulation results}
\label{model}
\begin{figure}[!h]
\centering
\epsfig{width=.38\textwidth,file=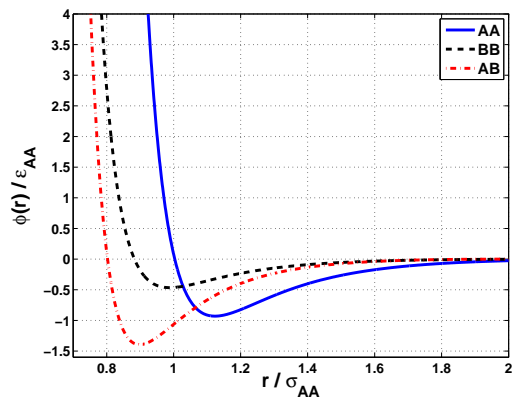}
\caption{Interaction potentials in dimensionless units.}
\label{fig1}
\end{figure}

A two-dimensional binary mixture consists of two kinds of particles $A$ and $B$.
The interatomic interactions are defined  by shifted and smoothed Lennard-Jones
potentials
\begin{equation}
\phi_{\alpha\beta}(r)=\left\{
\begin{array}{ll}
\phi_{\alpha,\beta}^{LJ}(r)+A_{\alpha\beta}+B_{\alpha\beta}r+C_{\alpha\beta}r^2
&\textrm{if $r\le R^{cut}_{\alpha\beta}$,}\\
0&\textrm{if $r> R^{cut}_{\alpha\beta}$,}
\end{array}\right.
\label{KA1}
\end{equation}
where
\begin{equation}
\phi_{\alpha\beta}^{LJ}(r)=4\epsilon_{\alpha\beta}
\bigg[\bigg(\frac{\sigma_{\alpha\beta}}{r}\bigg)^{12}
-\bigg(\frac{\sigma_{\alpha\beta}}{r}\bigg)^{6}\bigg].
\label{LJ}
\end{equation}
It is convenient to introduce reduced units, with $\sigma_{AA}$ being the unit of length and
$\epsilon_{AA}$ the unit of energy.
All the potentials given by Eq.~(\ref{KA1}) vanish with two zero derivatives at distances  $R^{cut}_{\alpha\beta}=2.5\sigma_{\alpha\beta}$. The parameters in Eq.~(\ref{LJ}) \cite{KA94} and in the smoothing part of Eq.~(\ref{KA1}) are given in Tab.~\ref{tab1}. The dependence of the potentials defined by Eq.~(\ref{KA1}) on the distance between particles is shown in Fig.~\ref{fig1}.

%%%%%%%%%%%%%%%%%%%%%%%%%%%%%
\begin{table}[!h]
\caption{Potential parameters.}
\begin{tabular}{|c|c|c|c|c|c|}
\hline
Particles&$\sigma_{\alpha\beta}$&$\epsilon_{\alpha\beta}$&$A_{\alpha\beta}$&$B_{\alpha\beta}$&$C_{\alpha\beta}$\\
\hline
AA&1.00&1.0&0.4527&-0.3100&0.0542\\
BB&0.88&0.5&0.2263&-0.1762&0.0350\\
AB&0.80&1.5&0.6790&-0.5814&0.1271\\
\hline
\end{tabular}
\label{tab1}
\end{table}
%%%%%%%%%%%%%%%%%%%%%%%%%%%%%%%

%%%%%%%%%%%%%%%%%%%%%%%%%%
%%%%%%%%%%%%%%%%%%%%%%%%%%

A composition of $A$ and $B$ particles that is stable in two-dimensions against
crystallization is chosen to be $65\%$ of particles $A$ and $35\%$ of particles $B$ \cite{BOPK09}.
For the one component system that is discussed below the potential of interaction is chosen to be
that of particles $A$.

\subsection{The perfect hexagonal structure}
\subsubsection{Finite temperature}

\begin{figure}
\centering
\epsfig{width=.38\textwidth,file=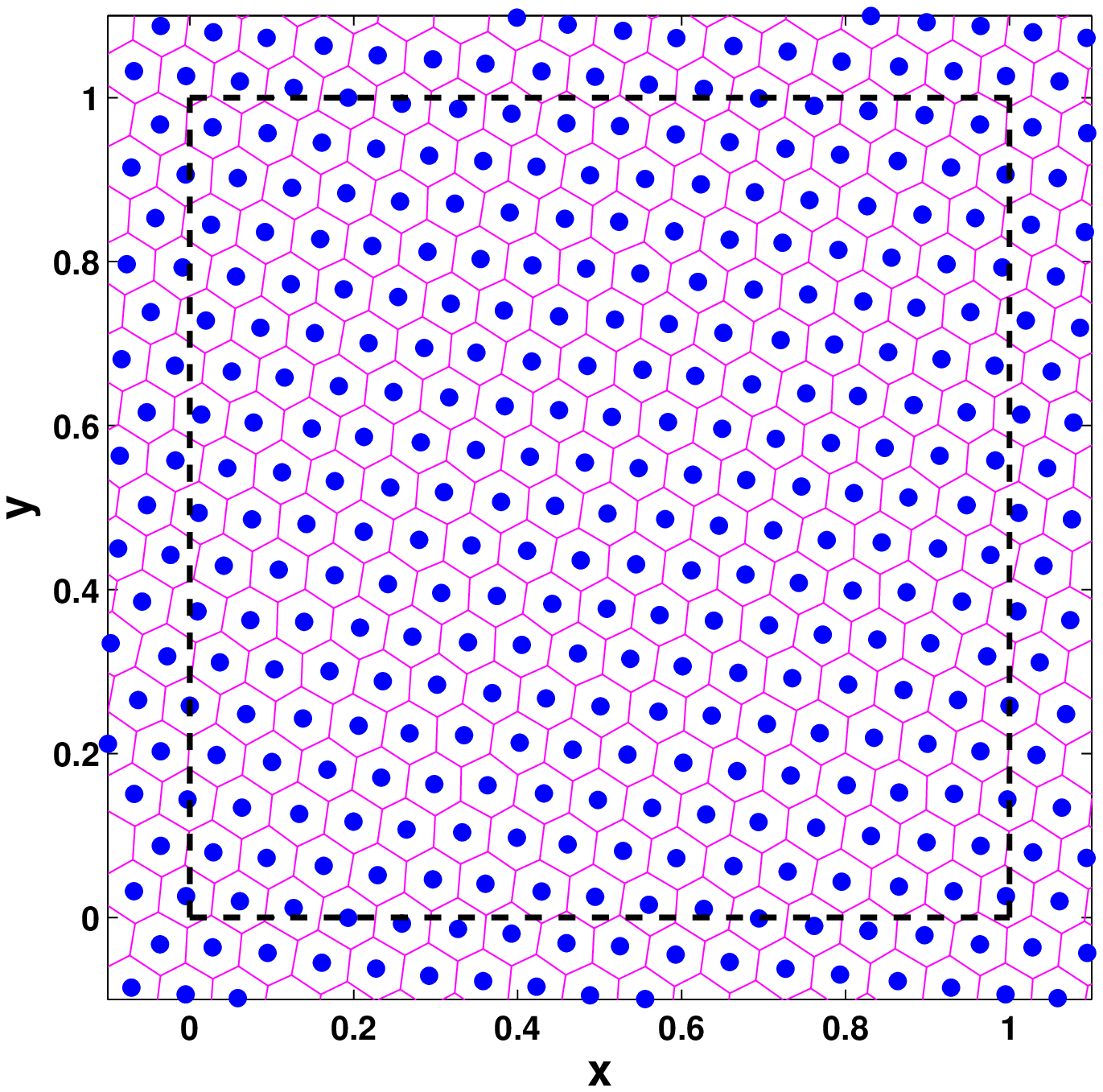}
\epsfig{width=.38\textwidth,file=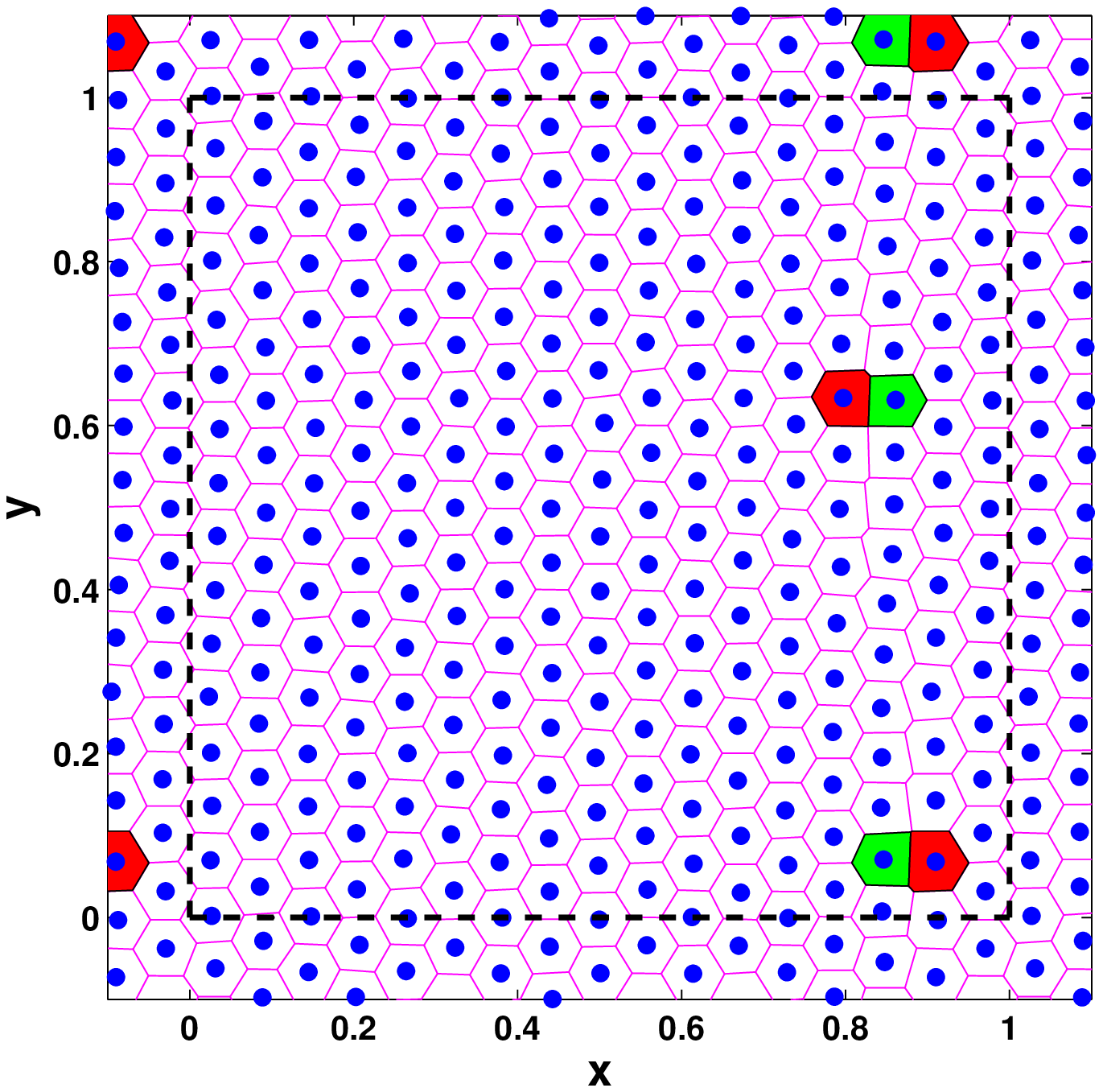}
\caption{Configurations of the one component system. Top panel - configuration {\bf I} with perfect hexagonal structure. Bottom panel - configuration {\bf II} with defects. The dotted line represents the simulation cell which is continued
periodically in both directions.}
\label{fig2}
\end{figure}

As a first step we studied the properties of a one component
system consisting of $N=256$ particles with the interaction potential of $A$ particles.
A Monte Carlo process
with $10^6$ sweeps at a chosen value of the shear stress was run using the shape-varying protocol described above.
We always begin our simulations from the liquid state, and cool down to a chosen temperature. This process
invariably leaves, even for a one component system, some defects in the self-forming crystalline hexagonal solid.
In other words, typically one finds, upon cooling, a configuration like the one shown in the lower panel
of Fig.~\ref{fig2}, denoted as configuration ${\bf II}$). These remaining defects can be
removed by straining the system back and forth as was done in Ref. \cite{ABHIMPS07}. The resulting perfect hexagonal
structure (the configuration ${\bf I}$) obtained in this way is shown in the
top panel of Fig.~\ref{fig2}.
%%%%%%%%%%%%%%%%%%%%%%%%%%
%%%%%%%%%%%%%%%
%%%%%%%%%%%%%%%%%%%%%%%%%%
\begin{figure}[!h]
\centering
\epsfig{width=.38\textwidth,file=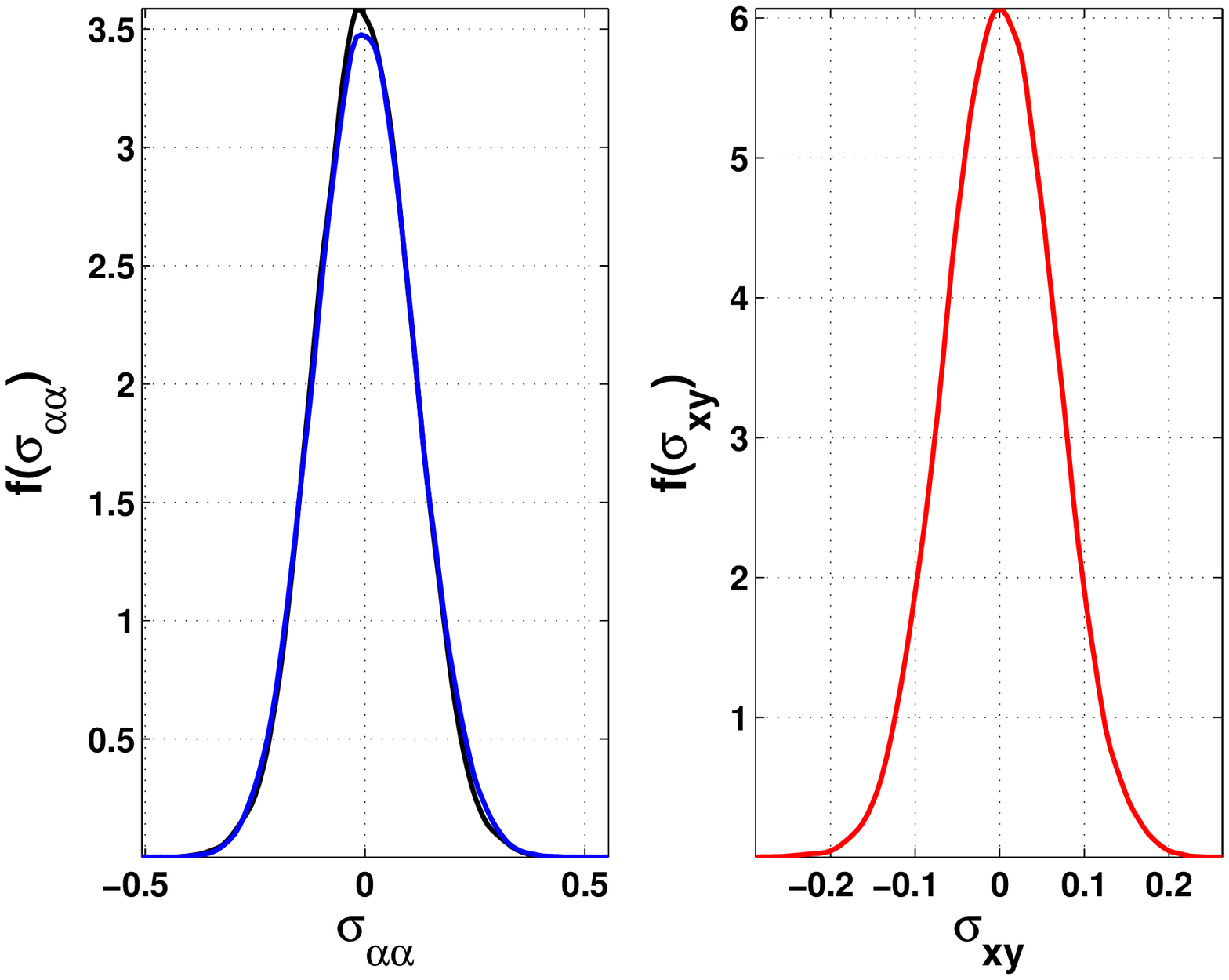}
\epsfig{width=.38\textwidth,file=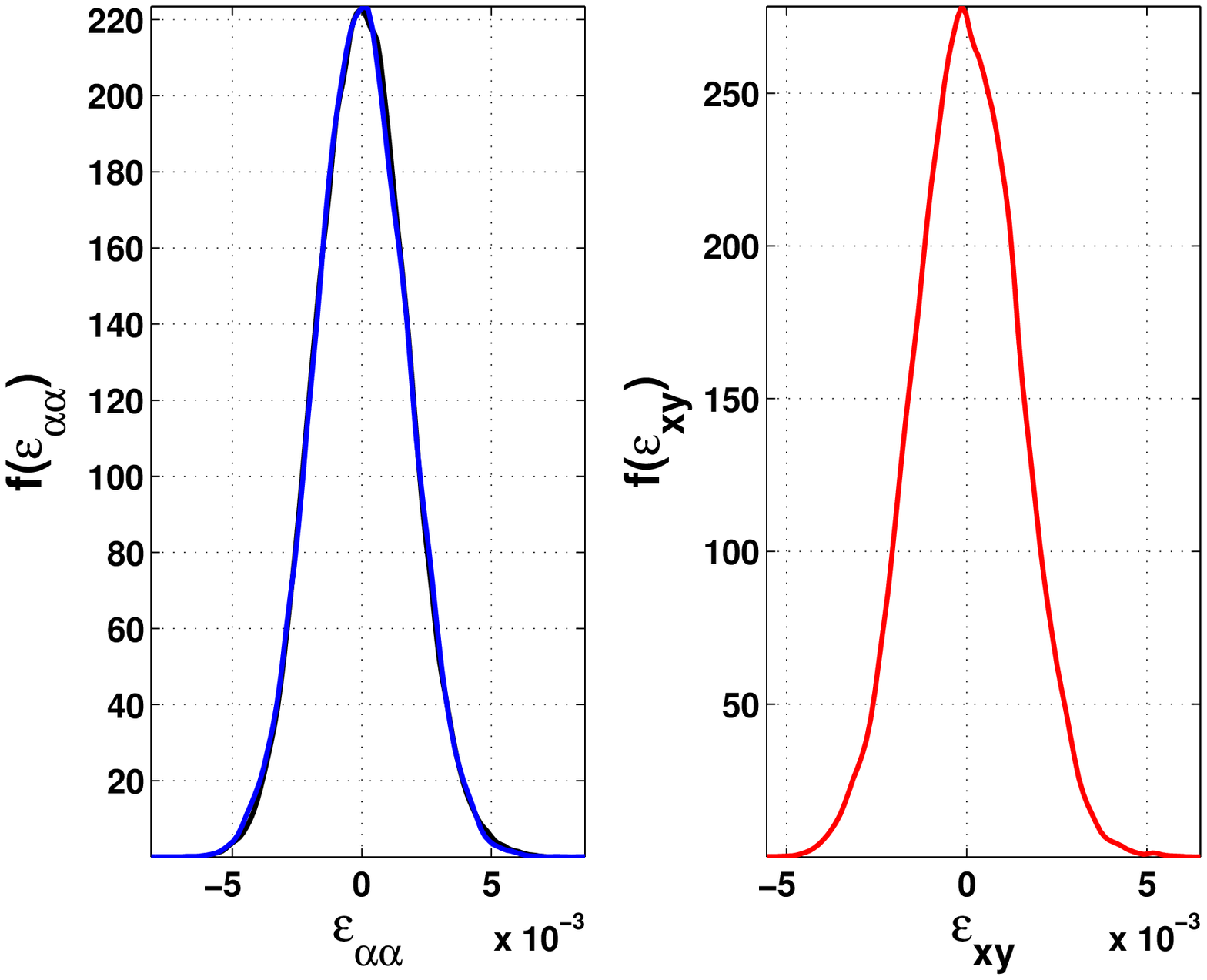}
\caption{Distributions of the stress tensor components (upper panel) and
strain tensor components (bottom panel) of the perfect hexagonal structure
at $\boldsymbol\sigma^{ext}=0 $ at temperature $T=0.05$ ($\alpha=x,y$).}
\label{fig3}
\end{figure}
%%%%%%%%%%%%%%%%%%%%%%%%%%

The distribution of the components of the internal stress and strain tensors
at zero pressure {\em and} with the external stress  $\B \sigma^{ext}=0 $ and $T=0.05$
is shown in Fig.~\ref{fig3}. The components of the internal stress tensor
are defined by
\begin{equation}
\sigma_{\alpha\beta}^{int}=\rho T \delta_{\alpha\beta}-
\frac{1}{2V}\sum\limits_{\B K,\B L}\sum\limits_{i\ne j}
\frac{\partial \phi_{\B K \B L}(r_{ij})}{\partial r_{ij}}\frac{r_{ij}^{\alpha}
r_{ij}^{\beta}}{r_{ij}},
\label{sigint}
\end{equation}
where $r_{ij}$ is the distance between particles $i$ and $j$, $\alpha,\beta=x,y$ denotes components of a
vector $\B r_{ij}$ and $\B K, \B L=A,B$ distinguish the kind of a particle. The strain tensor is defined here by
\begin{equation}
 \B \epsilon=\frac{1}{2}\big( \tilde{\B h_0}^{-1}\tilde{\B h}{\B h}{\B h_0^{-1}}-{\B I}\big),
 \label{StrEps}
\end{equation}
where $\B h_0=\langle \B h\rangle$.

%%%%%%%%%%%%%%%%%%%%%%%%%%
\begin{figure}[!h]
\centering
\epsfig{width=.38\textwidth,file=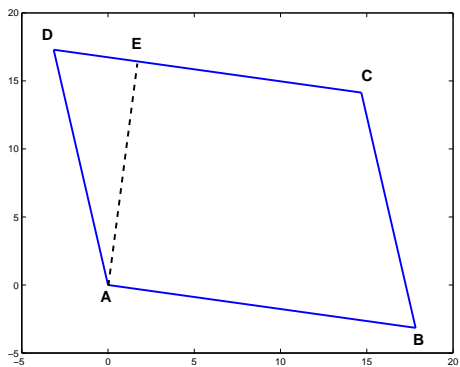}
\caption{Shape of the simulation box. The segment AE is perpendicular to the side CD}
\label{fig4}
\end{figure}
%%%%%%%%%%%%%%%%%%%%%%%%%%

For notational purposes it is more convenient to use a definition of shear deformation instead of Eq.~(\ref{StrEps}). A current shape
of the simulation box is shown in Fig.~\ref{fig4}. The strain (so-called engineer shear strain) is given by
\begin{equation}
 \gamma=\frac{L_{ED}}{L_{AE}},
 \label{EngStr}
 \end{equation}
where $L_{ij}$ is the distance between points $i$ and $j$. The same definition of strain is used in Eq.~(\ref{hG}). In order to define the deformation relative to a reference state we will use also the
quantity $\gamma_r=\gamma-\gamma_0$, where $\gamma_0=\langle\gamma\rangle_{\sigma^{ext}_{xy}=0}$.

\begin{figure}[!h]
\centering
\epsfig{width=.38\textwidth,file=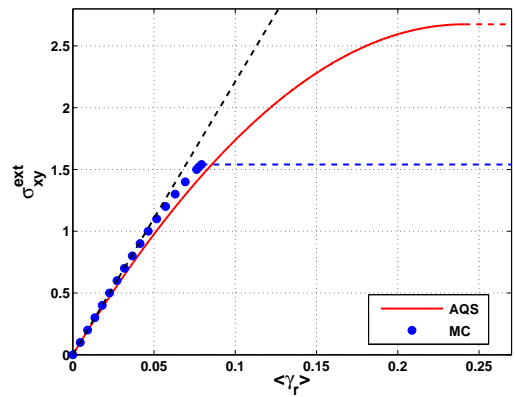}
\caption{Stress-strain dependence under stress control for system {\bf I}
(see Fig.~\ref{fig2}). The blue dots represent results of the Monte carlo simulations at $T=0.05$. The red line represents the prediction of the athermal quasistatic protocol (at $T=0$).
Note that at zero temperature the yield stress is considerably larger.}
\label{fig5}
\end{figure}

At this point the applied shear stress is increased in steps, and after each increase the
Monte Carlo process is run for 10 particles exchange sweeps, followed by a change in the shape $\B h$, followed again
by 10 particles sweeps and again a shape change, accumulating altogether to $10^6$ sweeps. As long as the chosen applied shear stress $\sigma_{xy}^{ext}$ is smaller
than $\sigma_{_{\rm Y}}\simeq 1.56$ the shear strain
 $\gamma_r$ reaches a constant mean value $\langle \gamma_r \rangle$ that does not change upon increasing the number of sweeps. When the applied shear stress $\sigma_{xy}^{ext}$
 exceeds $\sigma_{_{\rm Y}}$ the solid fails and the shear strain  grows without limit. This behavior is shown in Fig.~\ref{fig5}. It is noteworthy
 that the definition and the existence
of $\sigma_{_{\rm Y}}$ do {\bf not} depend on this step-wise increase in external shear stress. One could
go in one step to any value of the external shear stress and the response of the system will be the same, failing
only when $\sigma_{xy}^{ext}> \sigma_{_{\rm Y}}$.
%%%%%%%%%%%%%%%%%%%%%%%%%%
%%%%%%%%%%%%%%%%%%%%%%%%%
Note that in Fig.~\ref{fig5} the Monte Carlo results are compared with an AQS stress controlled protocol (see in Subsect. \ref{zeroT}
how this is defined and computed). One is not surprised that at zero temperature the yield stress is considerably
higher, and see below for more details. At this point it is enough to stress that $\sigma_{_{\rm Y}}$ depends on
temperature if one can wait. Only at zero temperature this quantity is absolute in the sense that no waiting time
is necessary for the system to fail. We return to this important issue in Sect. \ref{temp} where we estimate the
waiting time.

\begin{figure}[!h]
\centering
\epsfig{width=.38\textwidth,file=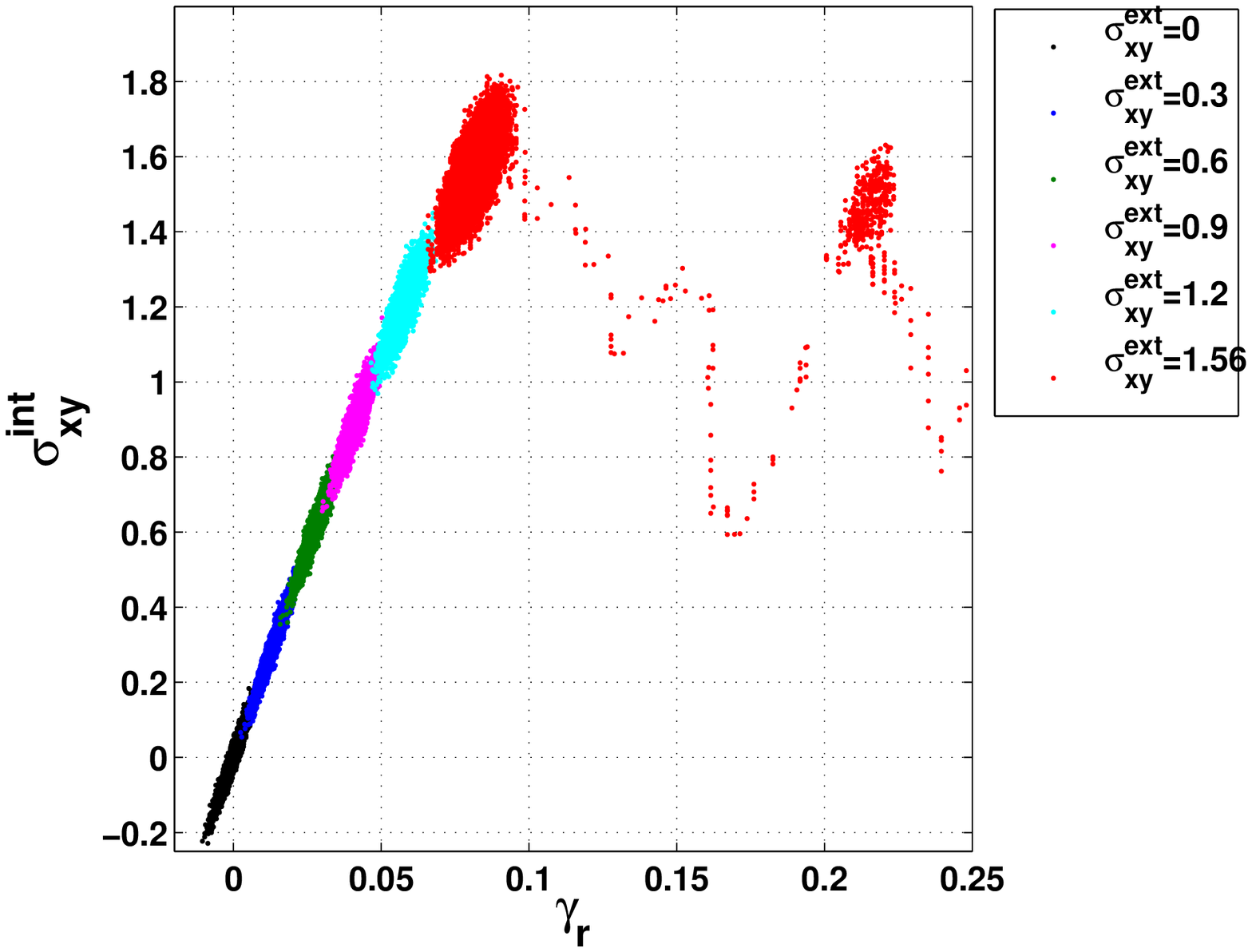}
\epsfig{width=.38\textwidth,file=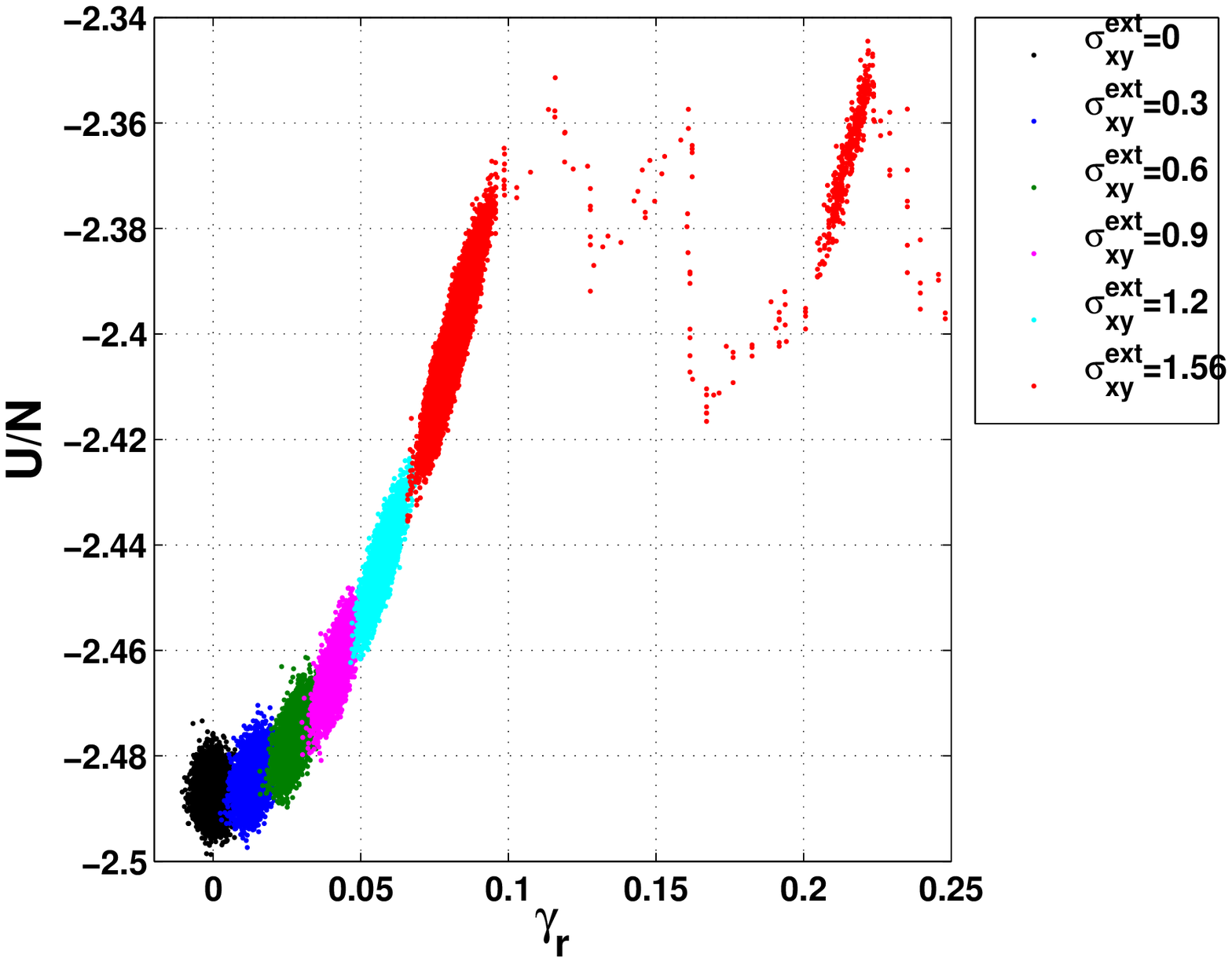}
\caption{The dependence of the internal stress (top panel) and the energy (bottom panel) on the strain under stress-control for
system {\bf I} (see Fig.~\ref{fig2}).}
\label{fig6}
\end{figure}

At finite temperature  the internal stress, the energy and the strain  fluctuate. The extent of these
fluctuations at $T=0.05$ is shown in
Fig.~\ref{fig6}. Below the yield stress the system exhibits elastic behavior.
When the yield stress is exceeded the system stays for a while in a series of
metastable states (each of which exhibiting ``elastic" behavior) whose life time becomes shorter and shorter
until the simulation box collapses entirely. Note that these metastable states are the elastic branches that
are seen very clearly in strain-controlled experiments, cf. Fig. \ref{fig11} and Fig.~\ref{fig15}. In that protocol the system
loses energy and releases strain upon reaching a saddle point bifurcation and lands on the next elastic branch
where it will stay forever if the strain does not increase. This is different from what is seen here, where
once $\sigma_{_{\rm Y}}$ is exceeded the system fails, even though it may reside for a while on metastable states.

%%%%%%%%%%%%%%%%%%%%%%%%%%%%%%%%%%%%%%%%%%%%%%%%%%%%%%%%%%%
\subsubsection{Zero temperature}
\label{zeroT}

\begin{figure}[!h]
\centering
\epsfig{width=.38\textwidth,file=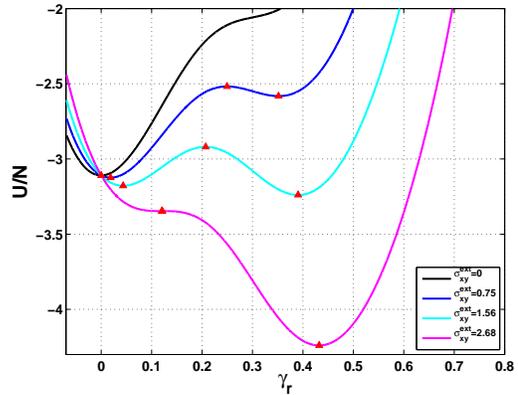}
\caption{ The strain dependence of the generalized enthalpy in the athermal
case for system $\bf I$. The input from the strain controlled experiment is the first curve at
$\sigma^{ext}_{xy}=0$. To this function we now add the term $-V \sigma^{ext}\gamma_r$ according to Eq.~(\ref{entGT})
to get all the other curves at varying values of $\sigma^{ext}_{xy}$.}
\label{fig7}
\end{figure}

In this subsection we show how to use the results of AQS strain control simulations
to predict the physics of AQS stress control loading.  Consider therefore the stress-strain
relation using the athermal limit in the NVT ensemble defined by
Eq.~(\ref{entGT}).
%%%%%%%%%%%%%%%%%%%%%%%%%%%%%%%%%%%%%%%%%%%%%
%%%%%%%%%%%%%%%%%%%%%%%%%%%%%%%%%%%%%%%%%%%
Imagine then that we run an AQS strain control simulation, and for every value of $\gamma$ we record
the energy $U(\{\B r_i\},\gamma)$ of the force free configuration after the non-affine relaxation took place.
In order to find the minimum of the function (\ref{entGT}) with regard to particle
positions and the strain at a given external stress we have to study the dependence
on strain of the generalized enthalpy. This dependence is shown in Fig.~\ref{fig7}. We reiterate that the contribution  $U(\{\B r_i\},\gamma)$ is independent of stress and is defined by minimizing the energy at given strain
via a relaxation of the particle positions.
%%%%%%%%%%%%%%%%%%%%%%%%%%%%%%%%%%%%%%%%%%%%%%
%%%%%%%%%%%%%%%%%%%%%%%%%%%%%%%%%%%%%%%%%%%%%%%%%%%%%%%%%%%%%%%
%%%%%%%%%%%%%%%%%%%%%%%%
\begin{figure}[!h]
\centering
\epsfig{width=.38\textwidth,file=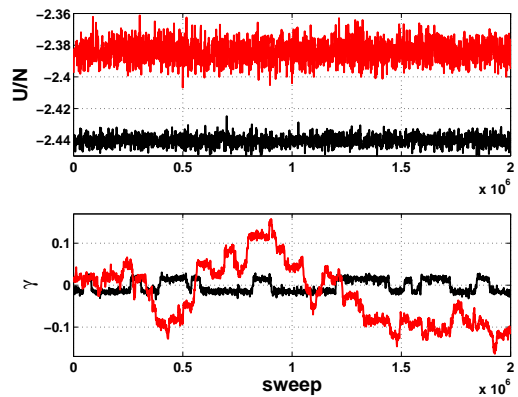}
\vskip 0.5 cm
\caption{ The evolution of the energy (top panel) and the strain $\gamma$ (botom panel) during the MC protocol
for system {\bf II} (see Fig.~\ref{fig2}) for two values of the temperature. The external stress $\B \sigma^{ext}_{xy}=0$, $T=0.05$ (black points) and $T=0.1$ (red points).}
\label{fig8}
\end{figure}
%%%%%%%%%%%%%%%%%%%%%%%%%%%%%%%%%%%%%%%%%%%
In the unstressed perfectly hexagonal structure there is only one minimum which is associated with a single reference state. Under applied stress there appears the metastable state
separated from the global minimum by a barrier. The barrier height decreases with increasing stress and it
disappears at the (zero-temperature) yield stress. We can now estimate the stress-strain relation from the
series of curves that are shown in Fig.~\ref{fig7}. As the stress increases the minimum of the curve
shifts to higher values of strain. The stress vs strain dependence that is read in this way is shown  in Fig.~\ref{fig5}. We see the almost perfect correspondence between the two curves for small values
of stress. The discrepancy
at higher strains results from having different ensembles: in the AQS protocol
the pressure varies non-monotonically with strain, in contrast to the Monte Carlo protocol
at constant pressure, which in the present case is $P=0$.

The increase in pressure in this AQS stress-controlled procedure eliminates the failure of the material that we
observe in the Monte Carlo stress-controlled protocol. Nevertheless we can predict the failure in the latter protocol
from the former. We need to focus on that value of the stress where for the first time the depth of the two minima
in Fig.~\ref{fig7} is the same. Note that this occurs at $\sigma=\sigma_{_{Y}}\approx 1.56$ in excellent agreement with the
results shown in Fig.~\ref{fig5}. Similar predictability will be shown below for the more complex examples.

%%%%%%%%%%%%%%%%%%%%%%%%
\begin{figure}[!h]
\centering
\epsfig{width=.38\textwidth,file=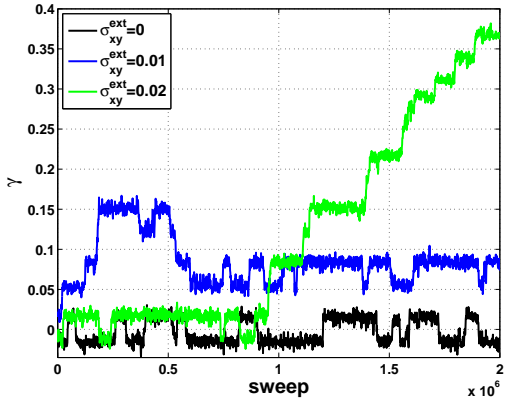}
\vskip 0.5 cm
\caption{Evolution of the strain $\gamma$ in the MC simulations under applied external stress.}
\label{fig9}
\end{figure}

\subsection{Hexagonal structure with defects}

\begin{figure}[!h]
\centering
\epsfig{width=.38\textwidth,file=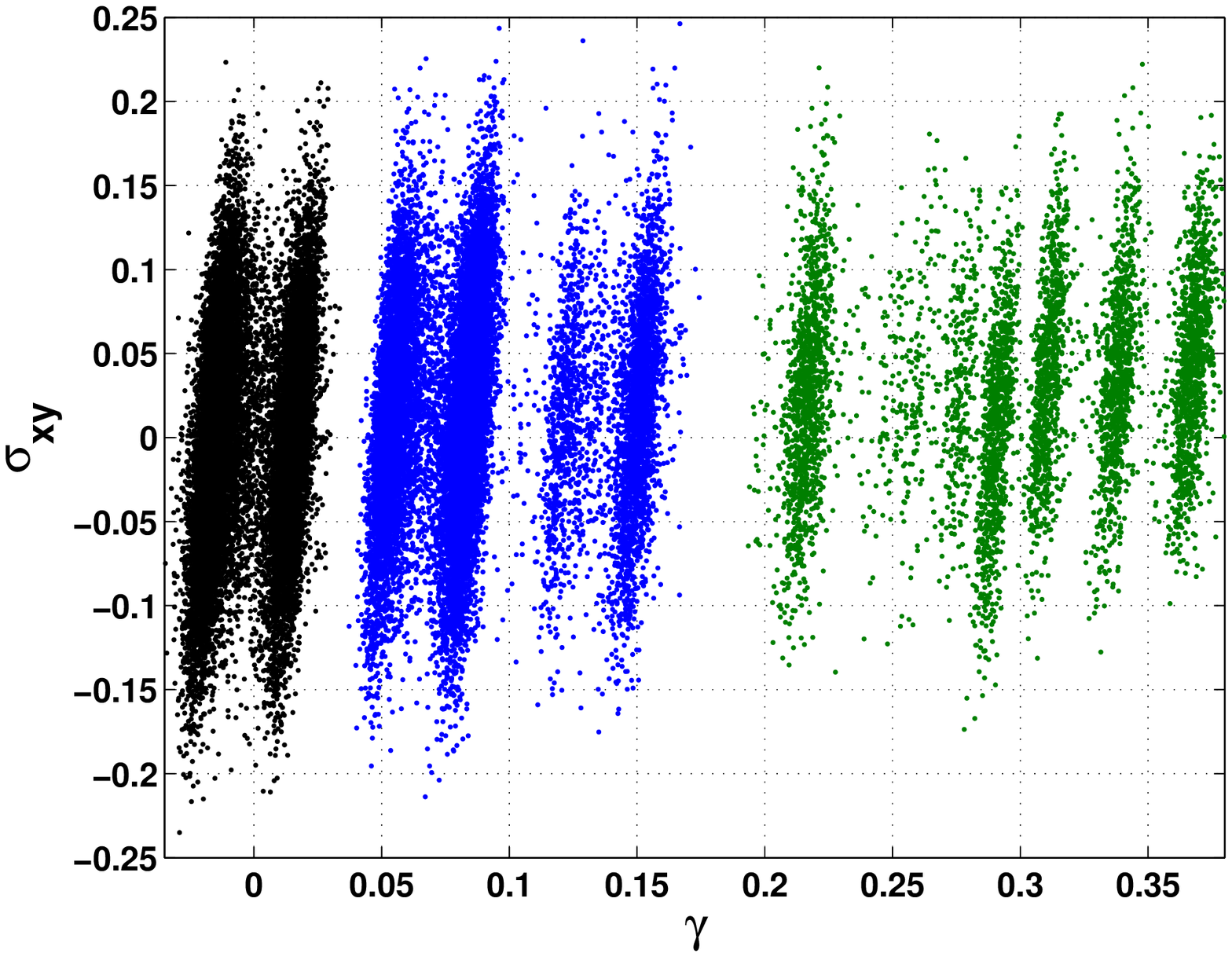}
\epsfig{width=.38\textwidth,file=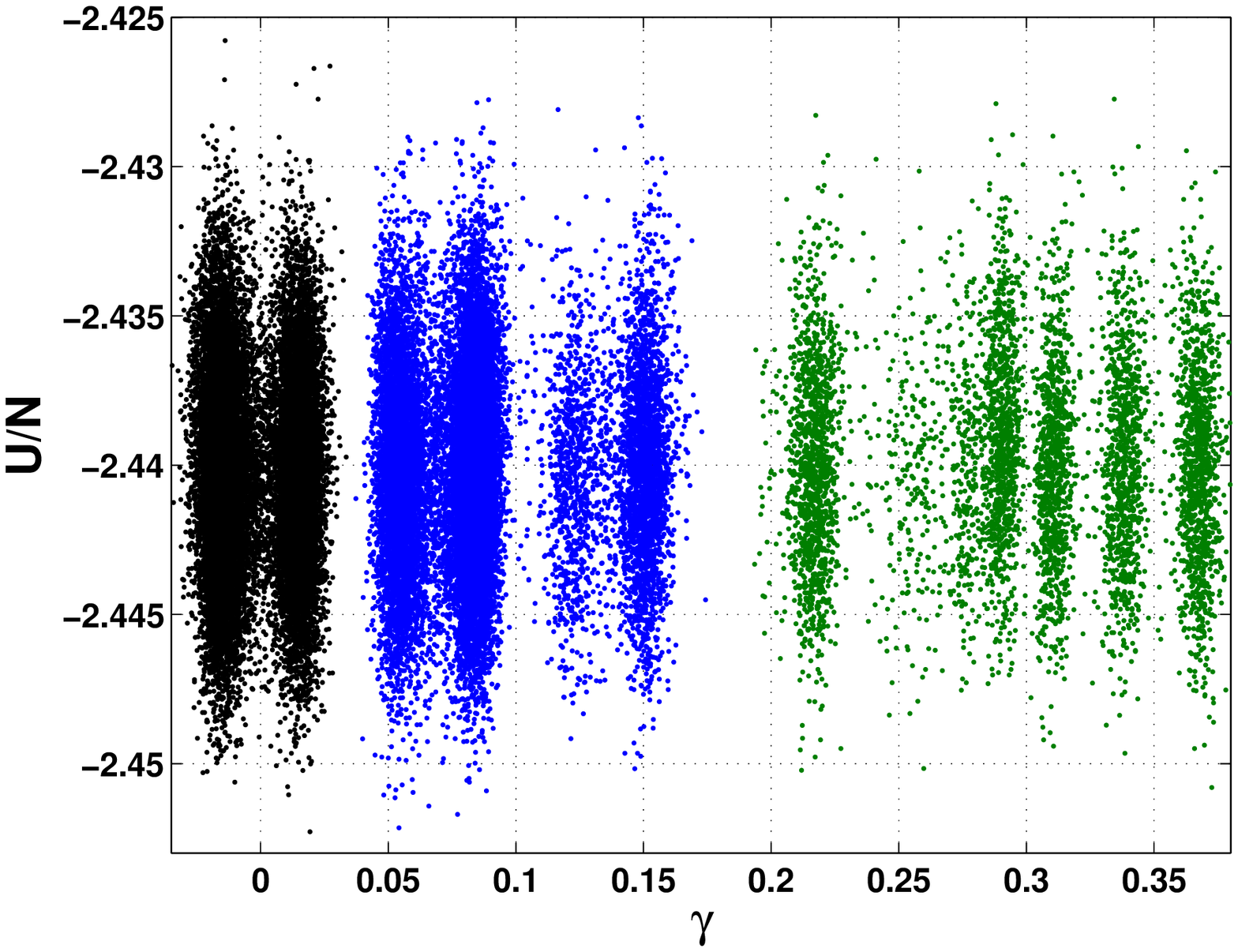}
\caption{ Dependence of the internal stress (top panel) and the energy (bottom panel) on the strain under stress-control for system {\bf II} (see Fig.~\ref{fig2}). Colors correspond to the legend in Fig.~\ref{fig9}}
\label{fig10}
\end{figure}

Our second system of interest is the hexagonal structure with a small number of defects whose concentration is about 2\%, as seen
in the lower panel of Fig. \ref{fig2}. Trajectories of measured values of the energy and the strain $\gamma$ as a function of the MC sweeps (here we used 2$\times 10^6$ sweeps) are shown in Fig.~\ref{fig8}. In contrast to the perfect hexagonal structure
the strain $\gamma$ displays at $T=0.05$ behavior typical to a bimodal distribution. Nevertheless, the comparison
with results for higher temperature $T=0.1$ which exhibit the liquid behavior (see also \cite{IMPS07}) shows that
the system at lower temperature is in a solid state. A few examples of the same dependence under applied external
stress are shown in Fig.~\ref{fig9}. One can see that at relatively small values of external stress there are
allowed transitions between available configurations. Then the applied stress exceeds some critical value
$ {\bf \sigma_Y}$ this dependence indicates  the permanent deformation of the simulation cell via a number of
metastable states (see Fig.~\ref{fig10}).

%%%%%%%%%%%%%%%%%%%%%%%%%%%%%%%%%%%%%%%%%%%%%
%%%%%%%%%%%%%%%%%%%%%%%%%%%%%%%%%%%%%%%%%%%

%%%%%%%%%%%%%%%%%%%%%%%%%%%%%%%%%%%%%%%%%%%%%
\begin{figure}[!h]
\centering
\epsfig{width=.38\textwidth,file=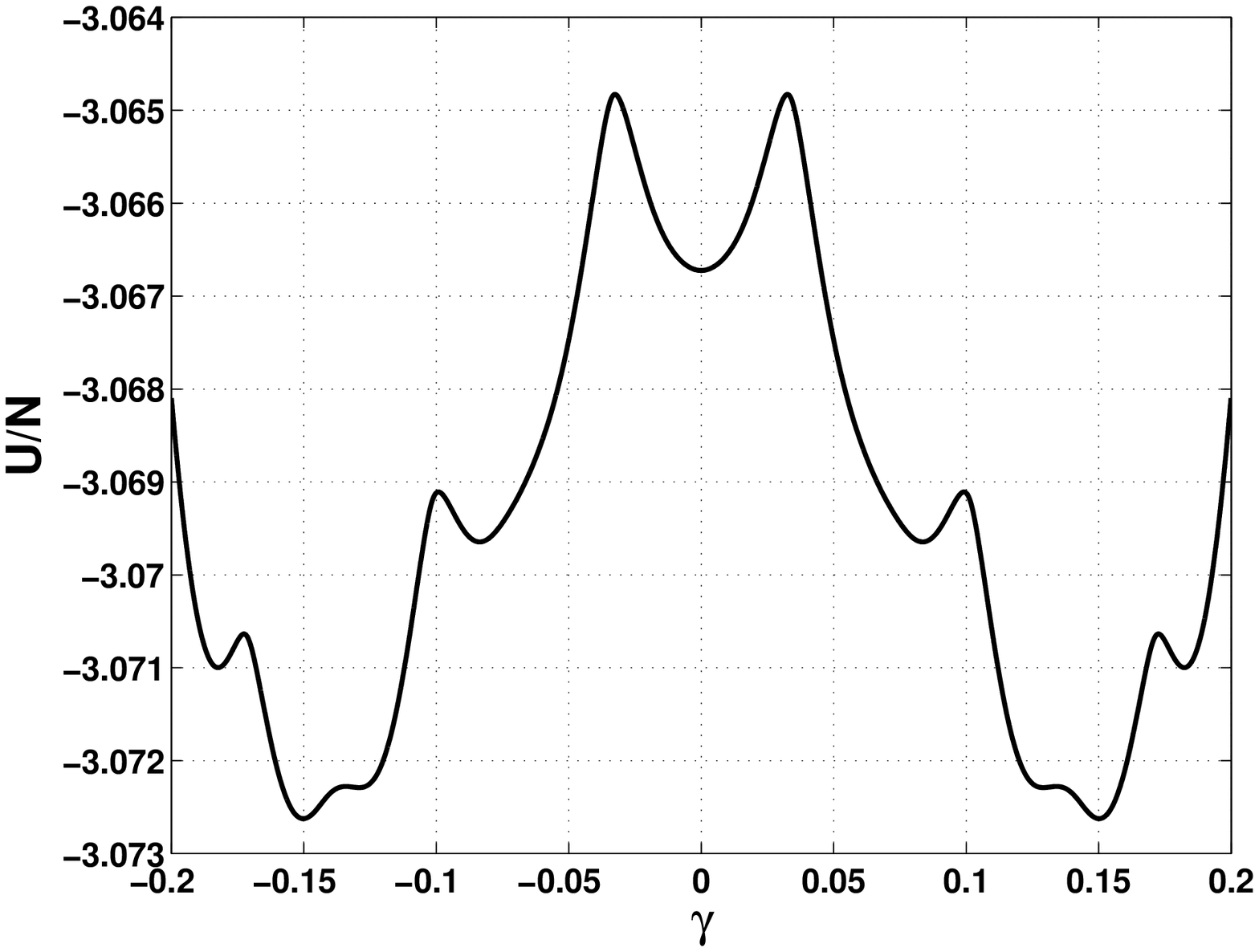}
\epsfig{width=.38\textwidth,file=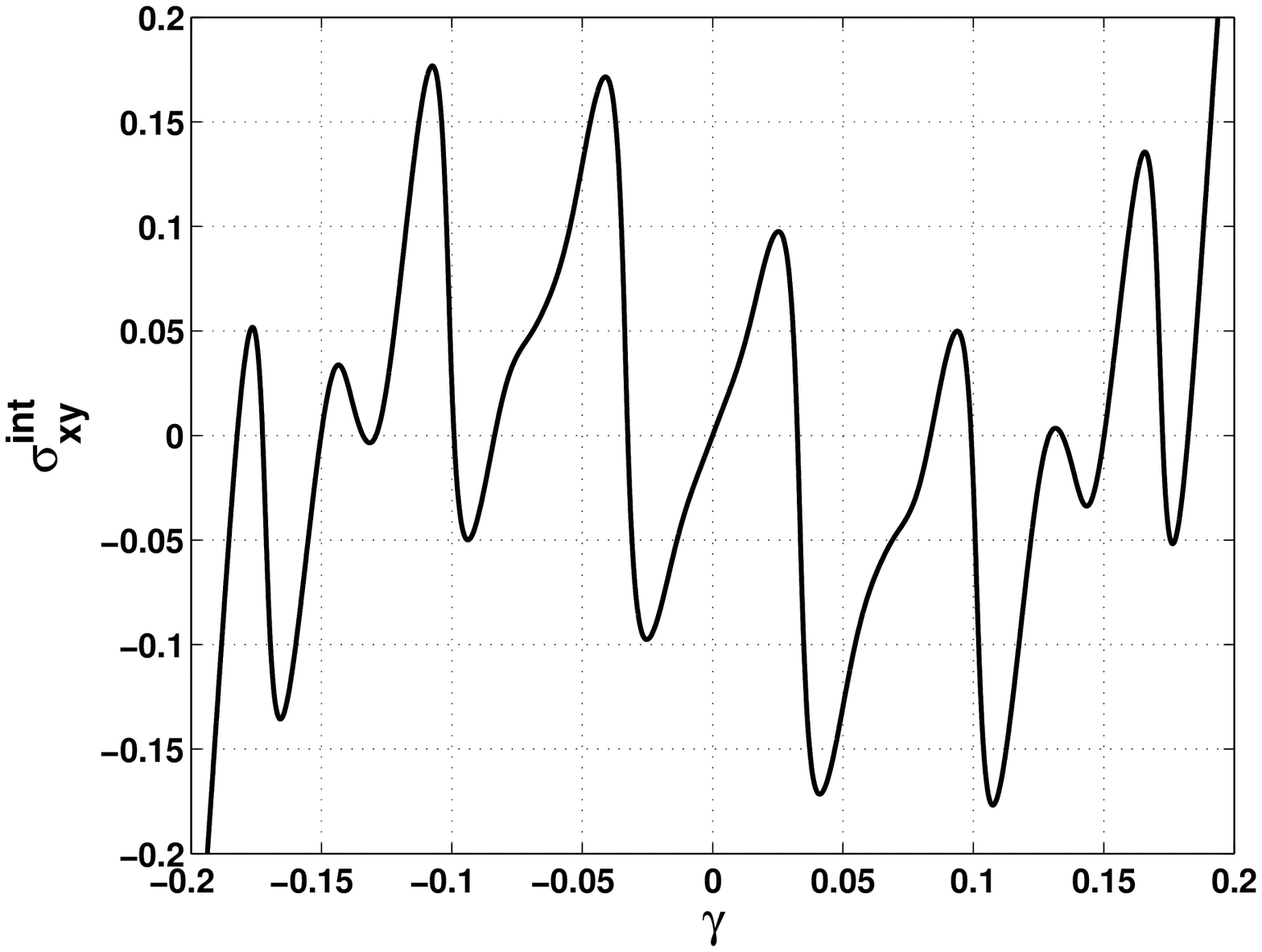}
\caption{ The simulation results in the frame of the AQS strain controlled protocol for the energy (top panel) and the internal stress (bottom panel).}
\label{fig11}
\end{figure}
%%%%%%%%%%%%%%%%%%%%%%%%%%%%%%%%%%%%%%%%%%%%%%%%%%%%%%%%%%%%%%%

The AQS strain controlled protocol (see Fig.~(\ref{fig11})) now reveals that $U(\{\B r_i\},\gamma)$ has a more complex landscape with a number of local minima.
The athermal analysis of the generalized enthalpy can be done again as explained above. The applied external stress shifts equilibrium positions similarly to
the MC results. Unfortunately, in the present case the amount of change in the pressure in the AQS protocol is too large to afford to say quantitative statement, leaving us with only a qualitative comparison.
\begin{figure}[!h]
\centering
\epsfig{width=.38\textwidth,file=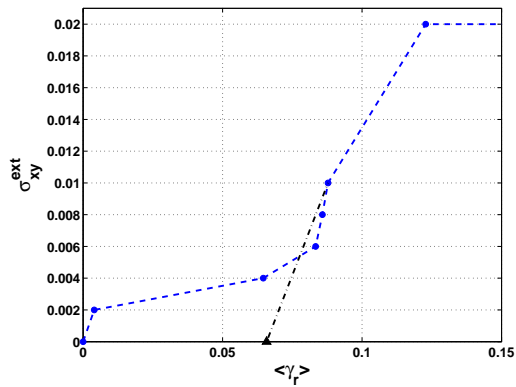}
\caption{Stress-strain dependence under stress control for system {\bf II}
(see Fig.~\ref{fig2}). In circles we show results of the Monte Carlo simulations with increasing value of $\sigma^{ext}_{xy}$. The black triangle represents the state
obtained by an MC simulation at $\sigma^{ext}_{xy}=0$ starting from an initial condition which is a configuration found by the Monte Carlo protocol at  $\sigma^{ext}_{xy}=0.01$.}
\label{fig12}
\end{figure}
%%%%%%%%%%%%%%%%%%%%%%%%%%%%%%%%%%%%%%%%%%%
Nevertheless, the simulation results show that transitions
between different minima can soften the material enormously, leading to a yield stress that is enormously smaller
than the corresponding one for the perfect hexagonal structure.

\subsection{The glass}

\begin{figure}
\centering
\epsfig{width=.38\textwidth,file=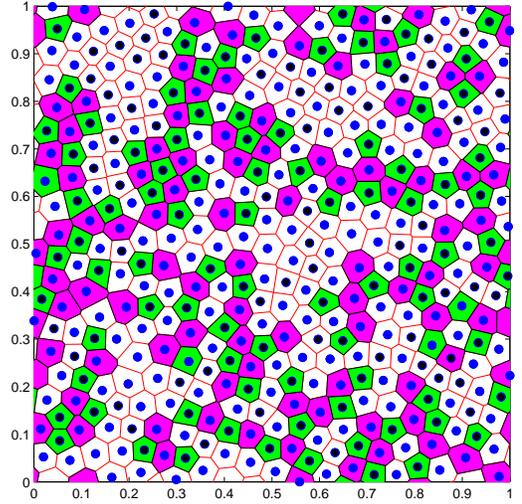}
\caption{The structure of the binary mixture. Blue circles corresponds to A particles, black circles to particles B .}
\label{fig13}
\end{figure}

In the glass simulations we employed 400 particles in the simulation cell. A typical configuration of the binary mixture which produces our model glass is shown in Fig.~\ref{fig13}.
The reader can already guess that the increased disorder seen in this figure will translate to an increased
complexity in the enthalpy landscape. Indeed, in Fig. \ref{fig15} we show the enthalpy landscape as computed
using the strain-controlled protocol and the changing landscapes upon the increase of the external stress.
%%%%%%%%%%%%%%%%%%%%%%%%%%%%%%%%%%%%%%%%%%%%%%%
%%%%%%%%%%%%%%%%%%%%%%%%%%%%%%%%%%%%%%%%%%%

%%%%%%%%%%%%%%%%%%%%%%%%%%%%%%%%%%%%%%%%%%%%%
\begin{figure}[!h]
\centering
\epsfig{width=.38\textwidth,file=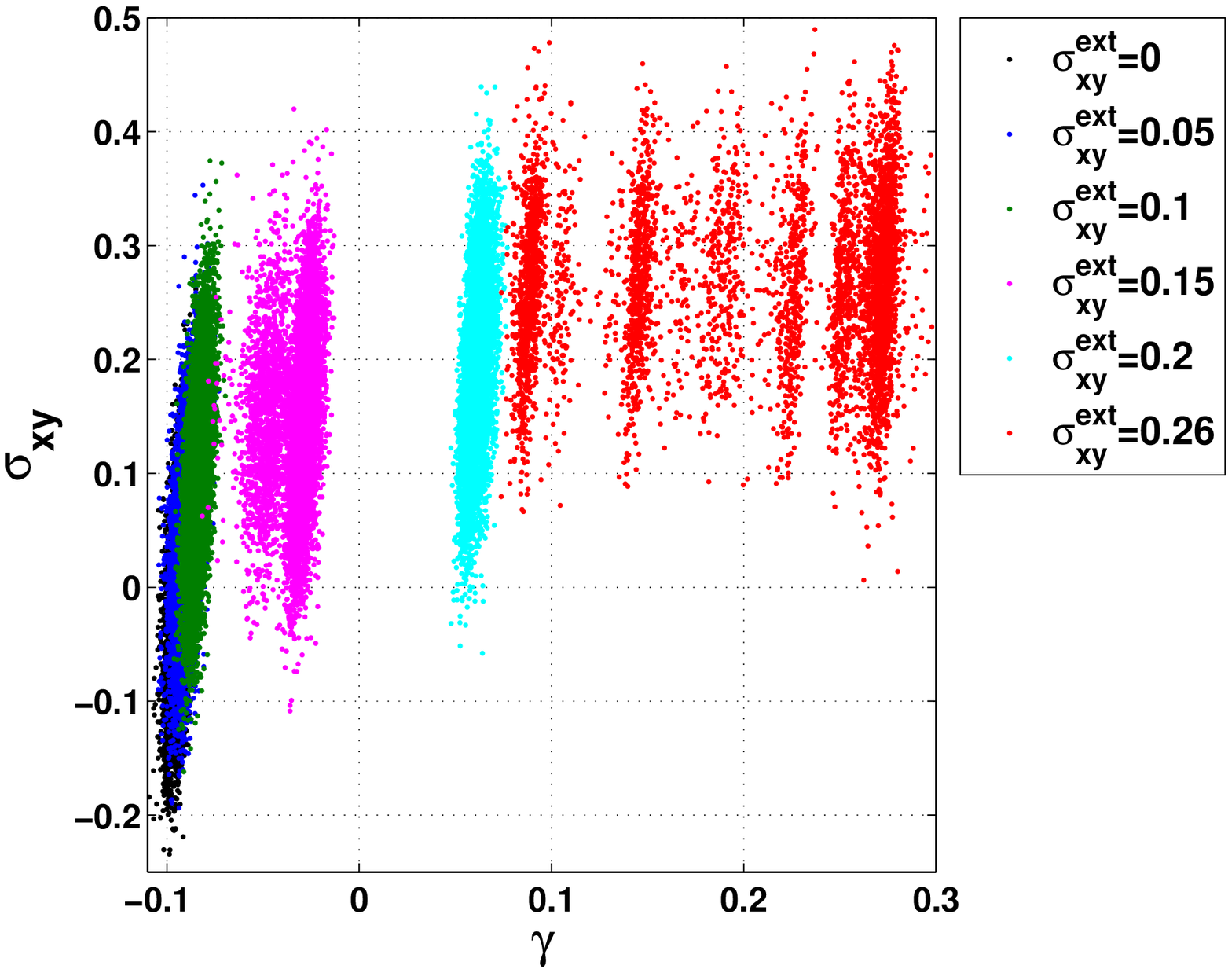}
\epsfig{width=.38\textwidth,file=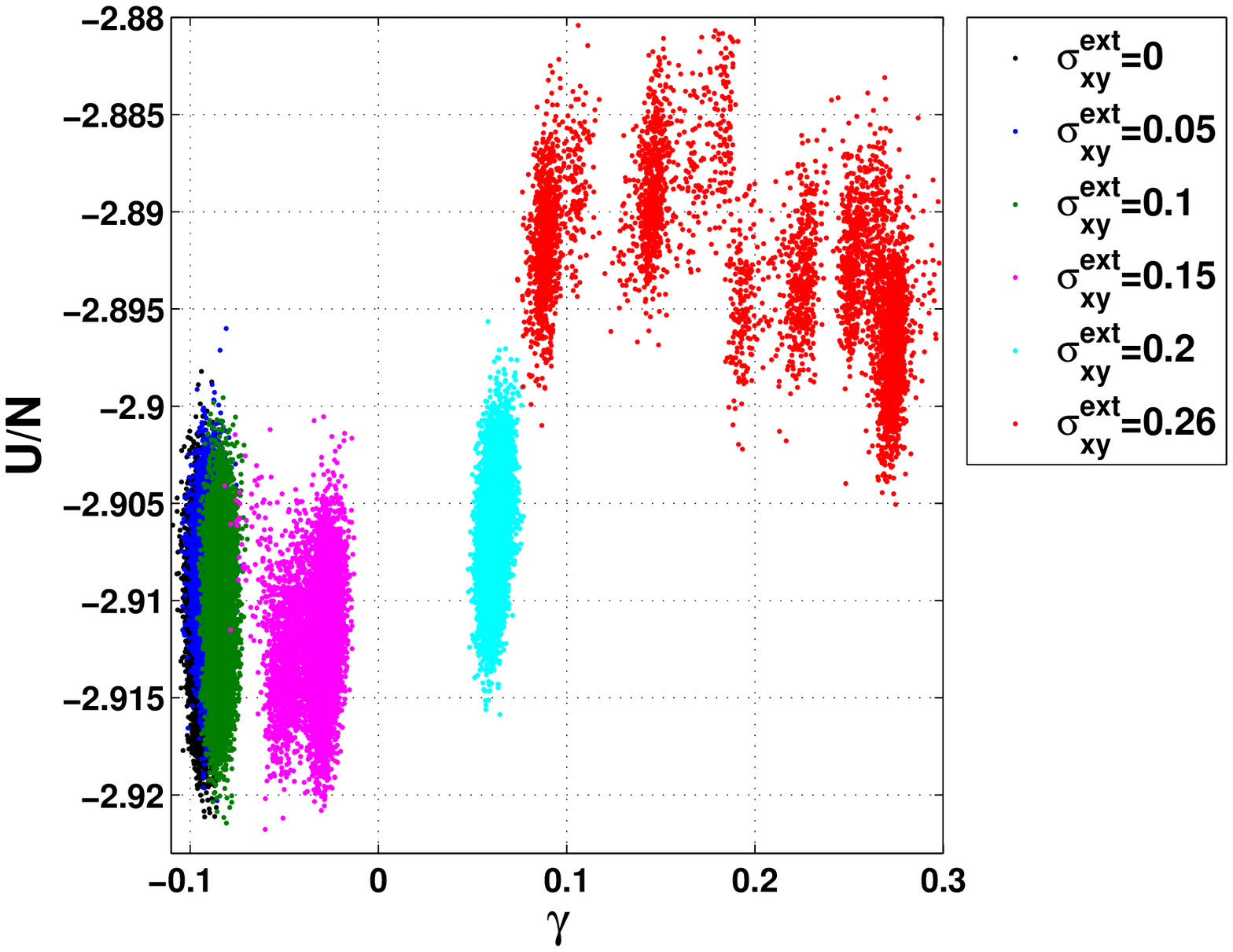}
\caption{ Dependence of the internal stress (top panel) and the energy (bottom panel) on strain under stress control for the glass model (see Fig.~\ref{fig13}).}
\label{fig14}
\end{figure}
%%%%%%%%%%%%%%%%%%%%%%%%%%%%%%%%%%%%%%%%%%%

\begin{figure}
\centering
\epsfig{width=.38\textwidth,file=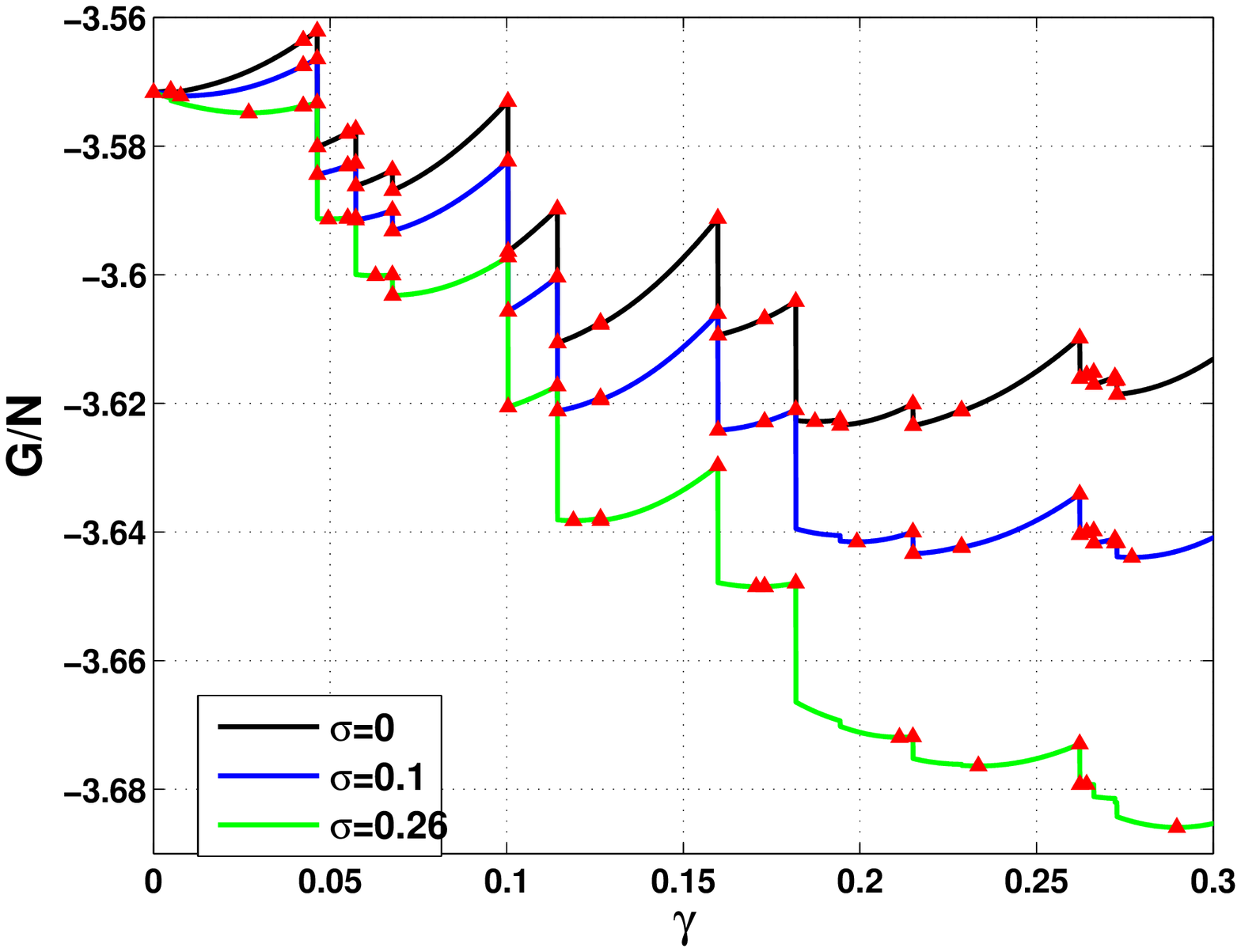}
\epsfig{width=.38\textwidth,file=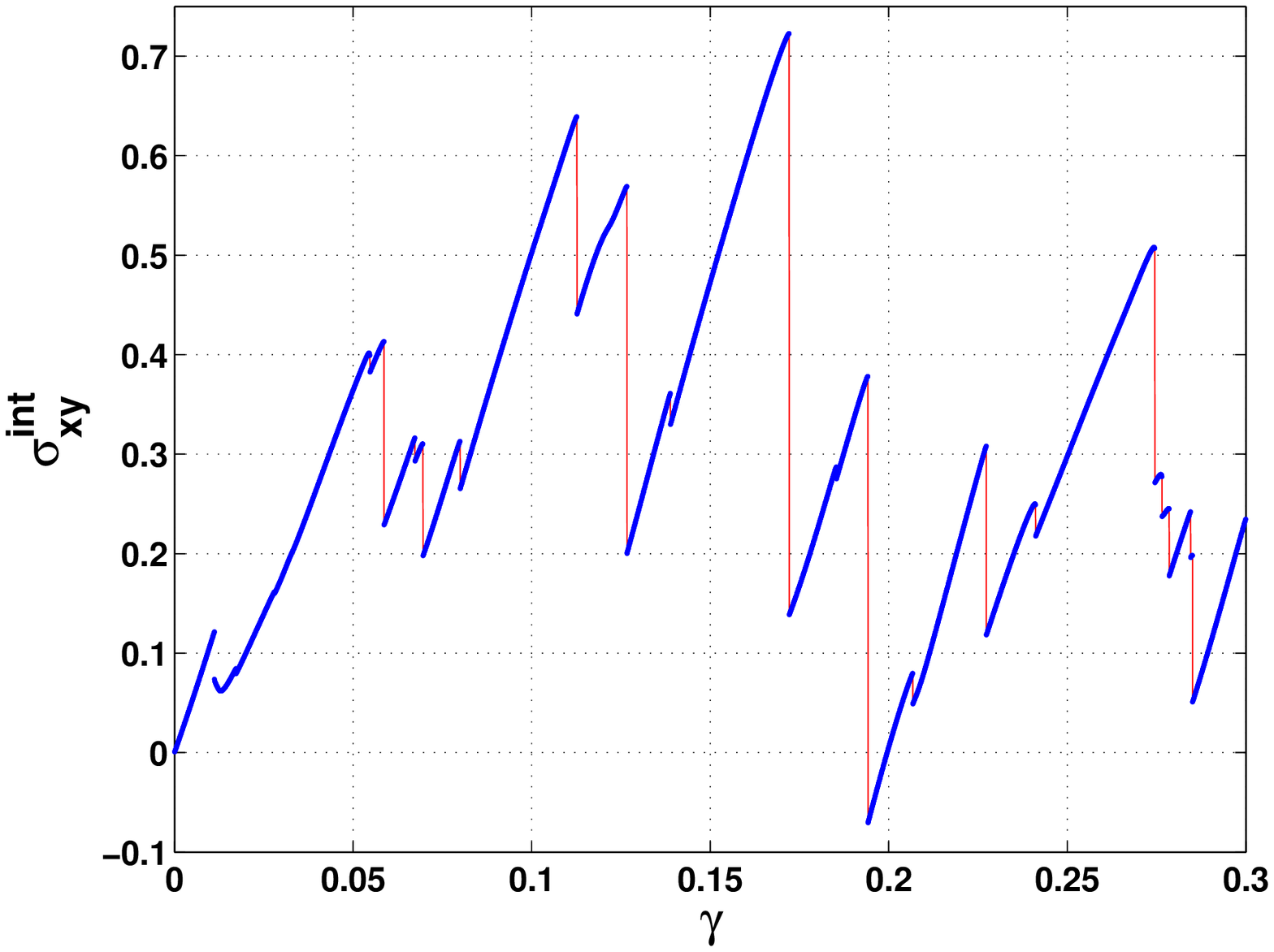}
\caption{Top panel -the strain dependence of the generalized enthalpy in the athermal
case for the glass. The input from the strain controlled experiment is the first curve at
$\sigma^{ext}_{xy}=0$. To this function we now add the term $-V \sigma^{ext}\gamma$ according to Eq.~(\ref{entG})
to get all the other curves at varying values of $\sigma^{ext}_{xy}$.
Bottom panel-stress vs strain in a strain controlled simulation of the response of the binary glass at AQS conditions.}
\label{fig15}
\end{figure}

\begin{figure}
\centering
\epsfig{width=.38\textwidth,file=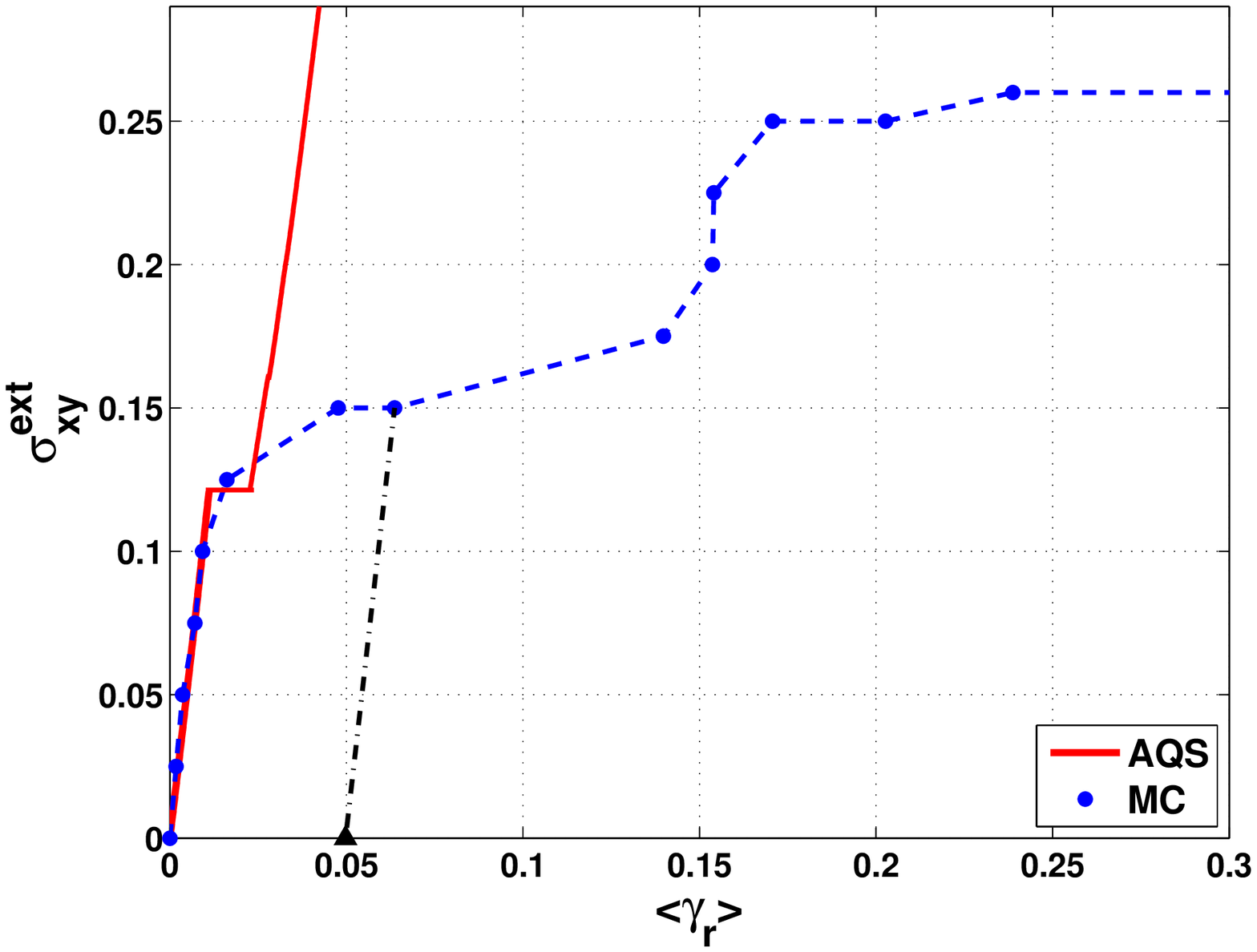}
\caption{Stress-strain dependence of the binary glass. We again stress that in the stress control simulation the external stress is fixed and shown
are the results for varying this fixed stress. In a triangle we show the state
obtained by the MC simulation at $\sigma^{ext}_{xy}=0$ with initial configuration from the run at  $\sigma^{ext}_{xy}=0.15$ }
\label{fig16}
\end{figure}

The corresponding results of the Monte Carlo simulation of the stress-strain dependence under stress control (in the glass simulations
we use 2$\times 10^6$ sweeps) are shown in Fig.~\ref{fig16}.
%%%%%%%%%%%%%%%%%%%%%%%%%%%%%%%%%%%%%%%%%%%%%%%
%%%%%%%%%%%%%%%%%%%%%%%%%%%%%%%%%%%%%%%%%%%
To understand these results we again turn to the strain control experiment at $T=0$, for which we exhibit the
stress vs strain trajectory in Fig. \ref{fig15}. Note again the immense difference between the two protocols: in strain
controlled simulations one sees many plastic instabilities, and in each of them the system releases a part of the stress
and a part of its mechanical energy. The strain is no longer a state variable due to the irreversible drops in energy.
In contrast, in the stress control experiment one is bound to get stuck at one of the elastic branches as long as the
stress is lower than the  yield stress $\sigma_{_{\rm Y}}$, which in the present case is about 0.26.
%%%%%%%%%%%%%%%%%%%%%%%%%%%%%%%%%%%%%%%%%%%%%%%

At zero temperature the stress control experiment can exhibit only one instability where the system fails, when
$\sigma^{ext}> \sigma_{_{\rm Y}}$. At finite temperatures one can observe multiple instabilities also in the stress control
protocol as the system overcomes
the barriers with the help of temperature fluctuations. Of course for a given external stress the waiting time will get longer and longer as the barrier
increases, until the barrier that is associate with the zero-temperature $\sigma_{_{\rm Y}}$ is reached. Finally note also the precise correspondence between
the zero-temperature and the finite temperature
trajectories for small values of stress in Fig.~\ref{fig15}. This correspondence can be maintained for much
higher values of stress and strain by reducing the temperature.

An important point to discuss is the fact that the glass is much more cohesive than structure {\bf II} even though
is has many more ``defects". The reason for this lies in the microscopic interactions that are exhibited in
Fig.~\ref{fig1}. We see there that the AB interaction is considerably deeper than the AA interaction, meaning that the
B particles act as pinning centers for the movement of A particles. This is in fact the deep reason why this mixture
is a good glass former. For the structure {\bf II} there is nothing that can pin the defects and they glide
happily under any minute strain or stress, which explains the low yield stress of that structure compared to the glass.
Indeed, this insight should be remembered whenever one wants to increase the cohesiveness of glasses, or to increase
their shear modulus or their yield stress. One should add particles that act effectively as pinning centers,
and see Ref. \cite{14Gen} for more details.
%%%%%%%%%%%%%%%%%%%%%%%%%%%%%%%%%%%%%%%%%%%%%%%%%%%%%%%%%%%
\section{Temperature Effects and Waiting Times}
\label{temp}

At this point we focus on values of the stress that are close to the yields stress
$\sigma_{_{\rm Y}}$, and particularly to the zero temperature value of this quantity (much of the discussion in this
section is however relevant for any instability point at lower values of stress). Stressing the system
at zero temperature will result in the system being stuck at a mean strain value $\langle \gamma \rangle$ that is less than the
value of the strain which is associated with the position of the highest barrier, denoted conveniently
as $\gamma_{_{\rm Y}}$. The question that we pose in this section is what is the waiting time $\tau$ (first passage time) for failure if the temperature is not zero. The problem is the classical one for escape over a barrier, but because
this is a saddle node bifurcation there are some special characteristics that need to be taken into account.

The general expectation for the waiting time is that it should scale like
\begin{equation}
\tau \sim \omega^{-1} \exp {\Delta G/T}  \ , \label{tau}
\end{equation}
where $\omega$ is the typical frequency of oscillations in the metastable minimum from which the system escapes,
and as before $\Delta G$ is the enthalpic barrier that becomes a saddle together with the minimum at $\sigma=\sigma_{_{\rm Y}}$. One knows that in a saddle node bifurcation the frequency $\omega\sim \sqrt{\lambda}$
where $\lambda$ is the lowest eigenvalue of the Hessian matrix. The latter goes to zero at the saddle bifurcation
like $\lambda\sim \sqrt{\gamma_{_{\rm Y}} - \gamma}$. As long as the harmonic approximation is relevant we can
therefore write
\begin{equation}
\omega \sim (\gamma_{_{\rm Y}} - \gamma)^{1/4} \ . \label{omeg}
\end{equation}
On the other hand the height of the barrier scales like
\begin{equation}
\Delta G \sim \lambda^3 \sim (\gamma_{_{\rm Y}} - \gamma)^{3/2} \ . \label{delG}
\end{equation}
Using these scaling estimates in Eq.~(\ref{tau}) we see that formally the waiting time diverges
both at $(\gamma_{_{\rm Y}} - \gamma) \to 0$ and at $(\gamma_{_{\rm Y}} - \gamma) \to \infty$ with
a minimum waiting time at a temperature dependent value $(\gamma_{_{\rm Y}} - \gamma) =(T/6)^{2/3}$.
In reality however for any finite temperature we lose the relevance of the harmonic approximation
in the limit $(\gamma_{_{\rm Y}} - \gamma) \to 0$, and we need to use the next, nonsingular, anharmonic
correction to $\omega$. Also in the other limit, when $(\gamma_{_{\rm Y}} - \gamma)$ becomes large, we lose
the relevance of the scaling law (\ref{delG}), destroying the singularity in this limit.  Thus in both
limits we predict a nonsingular waiting time. The conclusion is that a precise estimate of the waiting time
calls for molecular dynamics simulations that are beyond the scope of this paper.

\section{Summary and Concluding Remarks}
\label{summary}

The main aim of this paper was to introduce a reliable simulational approach to stress controlled loading
of systems at zero or finite temperatures. The method of variable shape appears stable and useful, and we
exemplified it for a perfect hexagonal structure, the same structure with a few defects, and a generic binary
glass in 2-dimensions. The main conclusion is that the type of mechanical instabilities encountered in stress-controlled
protocols is one and the same as those seen in strain-controlled protocols, i.e saddle node bifurcations in which the Hessian matrix becomes unstable, sending one of its eigenvalues to zero. As a result, the knowledge of the possible
instabilities that are found in AQS strain-controlled simulations is very helpful forr understanding what is
happening in stress-controlled simulations, even at finite temperature. Of course, the difference in the type
of ensemble is not unimportant, and deviations in the system response from one protocol to the other
are expected and also found. Nevertheless approximating the enthalpy landscape with the help of the AQS
stain controlled simulations is shown to be very useful in clarifying what one should expect in a finite
temperature stress-controlled protocol. Examples of this understanding were given for all the three examples
treated in this paper. Finally we examined the issue of the waiting time for yield in stress-controlled simulations, making full use of the identification of the instabilities as saddle-node bifurcations. The conclusion there was that one can see a decrease or increase in the waiting time as a function of the distance from the instability, but
that eventually the waiting time is not singular.

We reiterate that all our stress-controlled simulations here were performed for zero pressure. It would be interesting
in the future to follow up on the present study with stress-controlled simulations when different component
of the stress tensor are kept constant, to see how the response of the system depends on such details.
We hope to report on such simulations in forthcoming publications. In addition, and maybe even more importantly,
the present protocol allows a very precise study of the yielding process itself. This study will be reported elsewhere.

\acknowledgments
This work had been supported in part by an "ideas" grant STANPAS of the ERC. We thank Eran Bouchbinder for some
very useful discussions.

%%%%%%%%%%%%%%%%%%%%%%%%%%%%%%%%%%%%%%%%%%%

%%%%%%%%%%%%%%%%%%%%%%%%%%%%%%%%%%%%%%%%%%%%%
\end{document}